\documentclass[%
 reprint,
 superscriptaddress,
 amsmath,amssymb,
 aps,
 pra,
]{revtex4-2}

\usepackage{graphicx}
\usepackage{dcolumn}
\usepackage{bm}
\usepackage{xcolor}

\newcommand{\unit}[1]{\ \mathrm{#1}}

\begin{document}

\preprint{APS/123-QED}

\title{Study on electro-optic noise in crystalline coatings toward future gravitational wave detectors}

\author{Satoshi Tanioka}
  \email{stanioka@syr.edu}
\author{Daniel Vander-Hyde}%
\affiliation{%
 Department of Physics, Syracuse University, Syracuse, New York 13244, USA
}%

\author{Garrett D. Cole}
\affiliation{
 Thorlabs Crystalline Solutions, 114 East Haley Street, Suite G, Santa Barbara, California 93101, USA
}%

\author{Steven D. Penn}
\affiliation{
 Department of Physics, Hobart William Smith Colleges, 300 Pulteney Street, Geneva, New York 14456, USA
}%

\author{Stefan W. Ballmer}
\affiliation{%
 Department of Physics, Syracuse University, Syracuse, New York 13244, USA
}%


\date{\today}

\begin{abstract}

Thermal noise in high-reflectivity mirror coatings is a limiting factor in ground-based gravitational wave detectors.
Reducing this coating thermal noise improves the sensitivity of detectors and enriches the scientific outcome of observing runs.
Crystalline gallium arsenide and aluminum-alloyed gallium arsenide (referred to as AlGaAs) coatings are promising coating candidates for future upgrades of gravitational wave detectors because of their low coating thermal noise.
However, AlGaAs-based crystalline coatings may be susceptible to an electro-optic noise induced by fluctuations in an electric field.
We investigated the electro-optic effect in an AlGaAs coating by using a Fabry-Perot cavity, and concluded that the noise level is well below the sensitivity of current and planned gravitational-wave detectors.
\end{abstract}

\maketitle

\section{Introduction}

Direct detection of gravitational waves (GWs) by ground-based laser interferometric gravitational wave detectors (GWDs) has provided unique insight into the Universe~\cite{Abbott2016, Abbott2017, Abbott2021}.
In the current laser interferometric GWDs, km-scale Fabry-Perot arm cavities are used which employ test mass mirrors coated with high-reflectivity amorphous coatings~\cite{Degallaix2019,Granata2020}.

The sensitivity of current GWD such as advanced LIGO (aLIGO) is partially limited by thermal noise arising from amorphous silica and titania-doped tantala coatings at their most sensitive frequency band~\cite{Harry2006, Gras2018}.
Future GWDs are planned to employ low thermal noise coatings so that one can explore further into the Universe with improved sensitivity~\cite{Punturo_2010, Adhikari2020, CEHS, Srivastava2022}.
Therefore, development of low thermal noise mirror coatings plays an important role in the development of future GWDs.

Crystalline gallium arsenide ($\mathrm{GaAs}$) and aluminum-alloyed gallium arsenide ($\mathrm{Al}_{x}\mathrm{Ga}_{1-x}\mathrm{As}$) coatings (referred to as AlGaAs coatings), which have demonstrated low thermal noise, are one of the coating candidates for future GWDs~\cite{Cole2013, Penn2019}.
In addition to exhibiting low elastic losses, optical absorption and scatter in AlGaAs are also low~\cite{Cole2016, Winkler2021}.
Therefore, AlGaAs coatings have a potential to improve the performance of GWDs, resulting in fruitful scientific outcomes.
There is a coordinated research effort to realize AlGaAs coating mirrors in future upgraded GWDs~\cite{Chalermsongsak2016, Marchio2018, Koch2019}.

While crystalline AlGaAs coatings can reduce thermal noise, they may also be susceptible to coupling from fluctuations in the electric field.
Refractive indices of AlGaAs coatings vary in proportion to the electric field via the electro-optic (EO) effect~\cite{Namba1961, yariv}.
Fluctuations in the electric field couples to the cavity length fluctuations through the change in refractive indices of coatings, and can show up as noise in a GWD~\cite{Abernathy, Marty}.

In order to investigate the impact of the noise induced by the EO effect in AlGaAs coatings, we have developed an experimental setup using a Fabry-Perot cavity.
In this study, we focused on the coupling between the electric field normal to the mirror surface and the cavity length.
From this experiment, we estimated the noise level of the EO effect, which was well below the strain sensitivity of current and future proposed GWDs.
We conclude that the EO noise in AlGaAs coating will not be a limiting noise source in these systems.

\section{Theory of electro-optic effect}

When an electric field is applied to certain materials, the refractive indices vary depending on this field.
This effect is called the electro-optic (EO) effect.
In this section, we briefly review the theory of the EO effect.
More details can be seen in the references \cite{Namba1961,yariv}.

\begin{figure}[htbp]
\includegraphics[width=8.6cm]{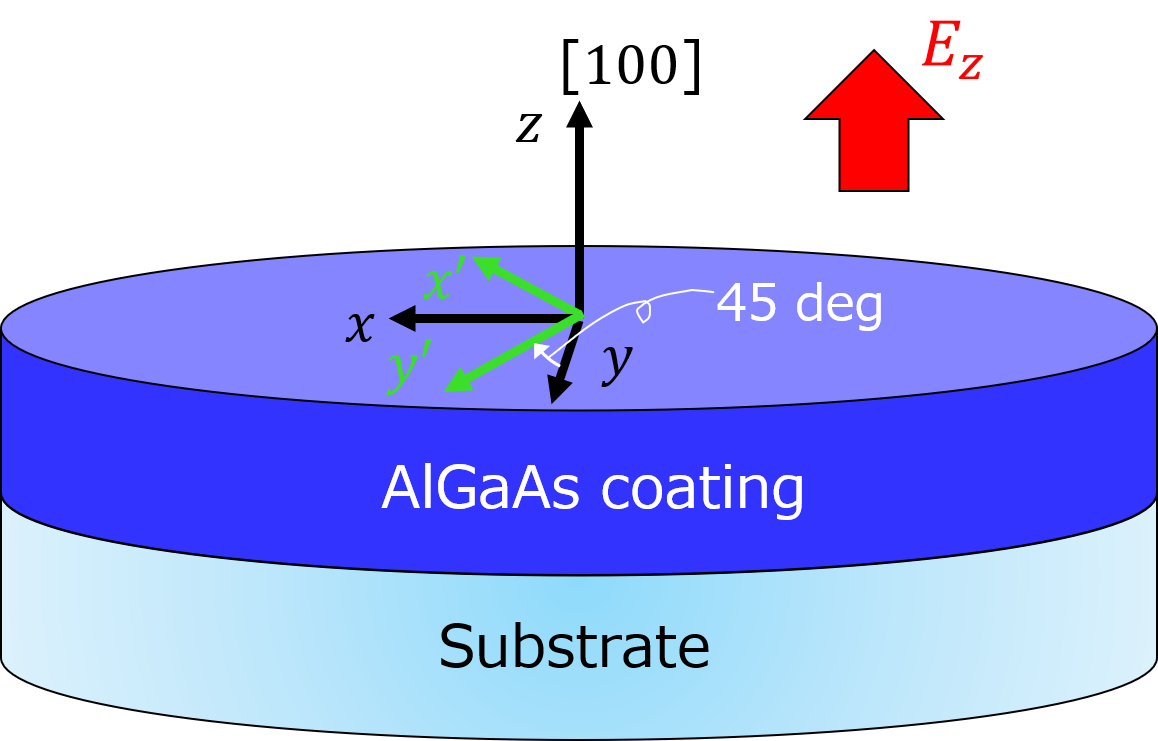}
\caption{
Schematic of the AlGaAs coating mirror.
The AlGaAs coating has the $[100]$ crystal axis normal to the surface.
$E_z$ represents the electric field along to $z$ axis.
}
\label{fig.coating}
\end{figure}

Refractive indices of a crystal can be expressed in terms of its index ellipsoid as
\begin{align}
    \frac{x^2}{n_x^2} + \frac{y^2}{n_y^2} + \frac{z^2}{n_z^2} = 1,
    \label{eq.index}
\end{align}
where $x$, $y$, and $z$ represent the coordinate axes, with the z-axis along the $[100]$ axis as shown in Fig.~\ref{fig.coating}.
And $n_x$, $n_y$, and $n_z$ are the three principal refractive indices with the crystallographic axes as the optical axes~\cite{yariv}.
For the case of zincblende crystals such as GaAs and AlGaAs, these refractive indices are $n_x=n_y=n_z=n_0$.

When the electric field is applied to the zincblende crystal, the index ellipsoid becomes~\cite{Namba1961,yariv}
\begin{align}
    \frac{x^2}{n_0^2} + \frac{y^2}{n_0^2} + \frac{z^2}{n_0^2} + 2r_{41}(E_{x}yz + E_{y}zx +E_{z}xy) = 1,
    \label{eq.refractive}
\end{align}
where $r_{41}$ represents the electro-optic coefficient.
If the electric field is applied along the $z$ axis, i.e., $E_x=E_y=0$, Eq. (\ref{eq.refractive}) becomes
\begin{align}
    \frac{x^2}{n_0^2} + \frac{y^2}{n_0^2} + \frac{z^2}{n_0^2} + 2r_{41}E_zxy = 1.
    \label{eq.determinant}
\end{align}
We define the new principal axes, $x'$, $y'$, and $z'$, when the electric field is applied as
\begin{align}
    \begin{pmatrix}
    x' \\ y' \\ z' \\
    \end{pmatrix}
    = \frac{1}{\sqrt{2}}
    \begin{pmatrix}
    1 & -1 & 0 \\ 1 & 1 & 0 \\ 0 & 0 & \sqrt{2}
    \end{pmatrix}
    \begin{pmatrix}
    x \\ y \\ z \\
    \end{pmatrix}.
\end{align}
By using the new coordinate, Eq.~(\ref{eq.determinant}) can be rewritten as
\begin{align}
     \left(\frac{1}{n_0^2}-r_{41}E_z\right)x'^2 + \left(\frac{1}{n_0^2}+r_{41}E_z\right)y'^2 + \frac{z'^2}{n_0^2} = 1.
\end{align}
Therefore, refractive indices of new principal axes, $n_{x'}$ and $n_{y'}$, become 
\begin{align}
    n_{x'} &= \left[\frac{1}{n_0^2}\left(1-n_0^2r_{41}E_z\right)\right]^{-1/2}, \\
    n_{y'} &= \left[\frac{1}{n_0^2}\left(1+n_0^2r_{41}E_z\right)\right]^{-1/2}.
\end{align}
Assuming that $n_0^2r_{41}E \ll 1$, these can be rewritten as
\begin{align}
    n_{x'} &= n_0 + \frac{1}{2}n_0^3r_{41}E_z, \\
    n_{y'} &= n_0 - \frac{1}{2}n_0^3r_{41}E_z.
\end{align}
Thus, an electric field changes the refractive indices of zincblende crystals such as GaAs and AlGaAs, hence AlGaAs coatings.
When the polarization of beams are aligned to new principal axes, $x'$ or $y'$, of AlGaAs coatings, optical path lengths in the coatings can be perturbed by the EO effect, causing perturbations in the phase of the reflected beam.
If the polarization is not aligned to $x'$ or $y'$ axes, the EO effect introduces birefringence.

It should be noted that $x'$, and $y'$ axes are $45$ degree rotated with respect to the positive $z$-axis as shown in Fig.~\ref{fig.coating}~\cite{Namba1961, yariv}.
For the case of GaAs and AlGaAs, $x$ and $y$ axes correspond to $[010]$ and $[001]$ directions, respectively.
Similarly, $x'$ and $y'$ axes are along the $[0\bar{1}1]$ and $[011]$.
Therefore, the changes of refractive indices due to the normal electric field are induced in principal axes of $[0\bar{1}1]$ and $[011]$ directions.

Crystalline AlGaAs coatings may be susceptible not only to the EO effect, but also to the piezoelectric effect~\cite{Nye1985}.
However, this effect does not directly couple to the cavity length fluctuations when the electric field is normal to the surface~\cite{Fricke1991}.
In this study, we only consider the EO effect that is much more dominant coupling source than the piezoelectric effect.

\section{Experiment}

\subsection{Setup}

\begin{figure}[htbp]
\includegraphics[width=8.6cm]{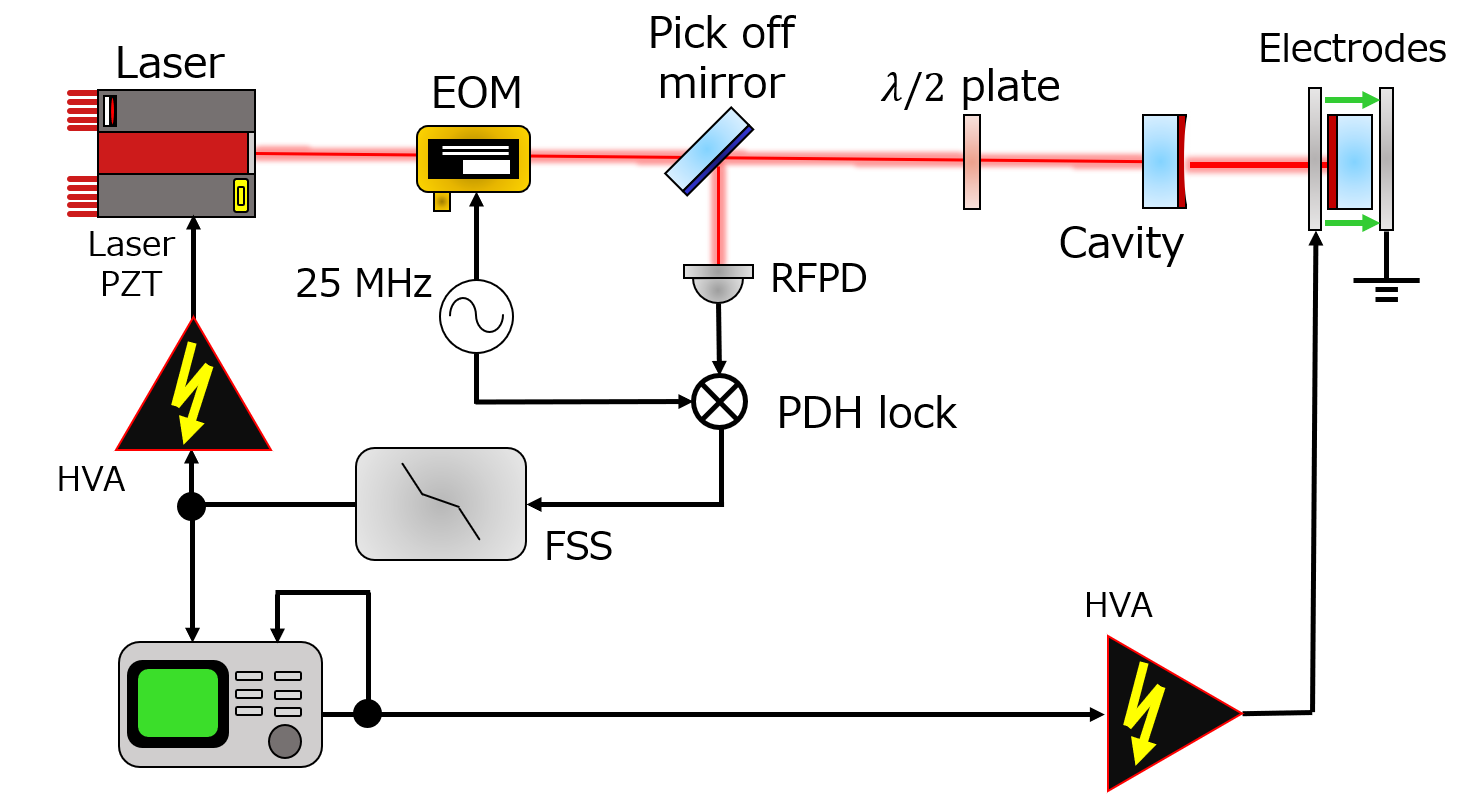}
\caption{
Schematic figure of experimental setup.
Laser frequency is locked to the cavity by PDH method. The reflected beam is detected by a radio frequency photo detector (RFPD) and then the signal is electrically demodulated. The demodulated signal is filtered by frequency stabilization servo (FSS) and then fed back to the laser PZT through a high-voltage amplifier (HVA).
The input mirror is an amorphous coating mirror which has a radius of curvature of $0.33\unit{m}$.
The end AlGaAs coating mirror, which is a flat mirror, is placed between two aluminum electrodes which apply the electric field normal to the mirror surface.
Voltage is applied to the front electrode through an HVA, and back electrode is grounded.
Polarization of the input beam is adjusted by a $\lambda/2$ plate.
}
\label{fig.setup}
\end{figure}

\begin{table}[htbp]
\caption{\label{tab.param}%
Parameters of experimental setup.}
\begin{ruledtabular}
\begin{tabular}{lcc}
Symbol & Description & Value \\
\colrule
$\lambda$ & Laser wavelength & $1064\unit{nm}$ \\
$L$ & Cavity length & $0.105\unit{m}$ \\
$x$ & Aluminum alloying fraction & 0.92 \\
$d_{\mathrm{H}}$ & Thickness of $\mathrm{GaAs}$ & $76.43\unit{nm}$ \\
$n_{\mathrm{H}}$ & Refractive index of $\mathrm{GaAs}$ & $3.48$ \\
$d_{\mathrm{L}}$ & Thickness of $\mathrm{Al_{0.92}Ga_{0.08}As}$ & $89.35\unit{nm}$ \\
$n_{\mathrm{L}}$ & Refractive index of $\mathrm{Al_{0.92}Ga_{0.08}As}$ & $2.98$ \\
\end{tabular}
\end{ruledtabular}
\end{table}

In order to experimentally investigate the EO effect in AlGaAs coatings, we developed an optical setup using a Fabry-Perot cavity.
Fig.~\ref{fig.setup} shows the schematic of the experimental setup.
The Fabry-Perot cavity is composed of two high-reflectivity mirrors --- an amorphous coating front mirror and AlGaAs coating end mirror.
The AlGaAs coating is composed of 35.5 periods (71 layers) of alternating $\mathrm{GaAs}$ and $\mathrm{Al_{0.92}Ga_{0.08}As}$, that have been transferred to a planar super-polished fused silica substrate.
The front mirror is curved mirror, and the end AlGaAs coating mirror has flat surface.
The finesse of the cavity is about $10^3$.

The laser frequency is locked to the cavity length by the Pound-Drever-Hall (PDH) technique~\cite{PDH}.
The extracted error signal is filtered by the frequency stabilization servo, and fed back to the PZT (piezo transducer) of an NPRO laser which actuates the laser frequency.
When the laser is locked to the cavity, fluctuations of the laser frequency, $\Delta\nu$ satisfies
\begin{align}
    \frac{\Delta \nu}{\nu} = -\frac{\Delta L}{L},
    \label{eq.cavity_fluctuation}
\end{align}
where $\Delta L$ is the cavity length fluctuations, and $\nu$ is the laser frequency.
The phase perturbation in AlGaAs coatings induced by the electric field is imprinted onto the cavity displacement, hence the PDH error signal.
By probing the PDH feedback signal, the displacement due to the EO effect can be measured.

\begin{figure}[htbp]
\includegraphics[width=8.6cm]{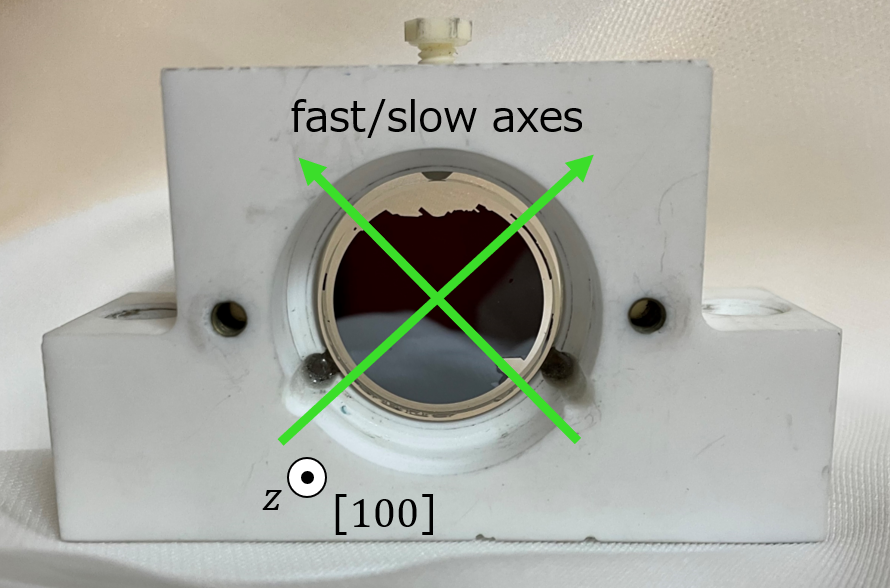}
\caption{
Front view of actual mirror mount for AlGaAs mirror without electrodes.
The AlGaAs mirror is clamped by a nylon screw with moderate torque.
The green arrows indicate the fast or slow axis where the refractive index is disturbed by the EO effect. The visible defects near the edges of the coating are due to excessive handling and are not typical of AlGaAs coatings. These defects do not impact the EO effect nor any results of this study.
}
\label{fig.mount}
\end{figure}

The input beam is linearly polarized, and its polarization can be aligned to the crystal axes of AlGaAs coatings by rotating a $\lambda/2$ plate in front of the Fabry-Perot cavity.
The AlGaAs sample mirror is installed as shown in Fig.~\ref{fig.mount}.

It is important to mention that AlGaAs coatings show larger birefringence than amorphous coatings.
AlGaAs coatings have two orthogonal distinct axes --- fast and slow axes which are aligned to $[0\bar{1}1]$ and $[011]$ orientations, respectively~\cite{thorlabs,Winkler2021}.
When the polarization of the beam is not aligned to the fast or slow axis, two distinct split resonant peaks can be generated as reported in previous works~\cite{Cole2013, Cole2016}.
No resonant peak splitting due to the birefringence was observed with our cavity used to measure the EO effect whose full-width-hald-maximum (FWHM) line-width is about $1.2\unit{MHz}$.
However, as described in the next section, two orthogonal polarization eigenmodes separated by about $500\unit{kHz}$ were observed when we replaced the input mirror to the one with higher reflectivity.
The AlGaAs fast and slow axes were identified by utilizing those split peaks.
The green arrows shown in Fig.~\ref{fig.mount} indicate the fast and slow axes of the AlGaAs coating.

The AlGaAs coated mirror and two electrodes are housed in the same mirror mount made of machinable glass, MACOR~\cite{MACOR}.
Each electrode has a hole with $3\unit{mm}$ diameter to pass the beam through.
The distances between the mirror surface and front electrode and back electrode are $0.39\unit{mm}$ and $0.20\unit{mm}$, respectively.
Source voltage is amplified to the front electrode by a HVA up to $2\unit{kV}$ peak to peak.
On the other hand, the back electrode is grounded, which introduces an electric field normal to the AlGaAs sample mirror surface.

\subsection{Axis identification}

\begin{figure}[htbp]
\includegraphics[width=8.6cm]{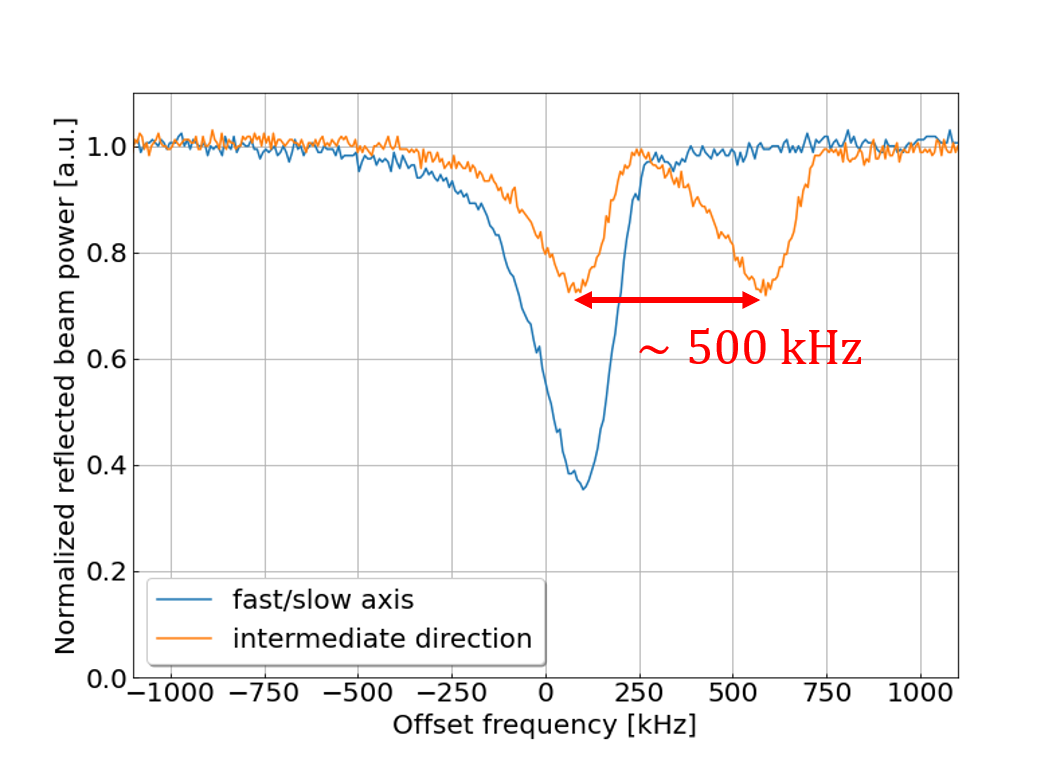}
\caption{
Response of the reflected beam power when the laser frequency is scanned.
As long as the input beam polarization is aligned to the fast or slow axis, the cavity shows single eigenmode as shown in blue curve.
On the other hand, when the polarization is misaligned from the fast or slow axis, two separated eigenmodes are observed due to the birefrincence in the AlGaAs coating (orange curve).
}
\label{fig.split}
\end{figure}

As described in the previous section, AlGaAs coatings have the fast and slow axes i.e., $[0\bar{1}1]$ and $[011]$ orientations, whose refractive indices are perturbed by the EO effect.
Prior to the measurements of the EO effect, we identified the fast and slow axes of the AlGaAs coating.
In order to determine the fast and slow axes, we used the higher-reflectivity mirror as the input mirror instead of the one used for the EO measurement.
With this configuration, the finesse of the cavity increased to about $4.5\times10^3$, and the FWHM line-width was about $300\unit{kHz}$.

Fig.~\ref{fig.split} shows the response of the reflected beam power when the laser frequency is scanned.
The laser frequency was swept by actuating the laser PZT with a triangle wave at $100\unit{Hz}$.
Then we adjusted the $\lambda/2$ plate to maximize or minimize the amount of the split peak.
When the beam polarization was aligned to the fast or slow axis, only single eigenmode was observed as shown by the blue curve.
Thus, we determined the angle of $\lambda/2$ plate which can align the input beam polarization to the fast or slow axis.
By tilting the $\lambda/2$ plate, the distinct split peak appeared as indicated by the orange line.
In our case, the split frequency of these two eigenmodes was about $500\unit{kHz}$.

After identifying the fast and slow axes, we switched the input mirror to what we originally used.
The reason why we employed the lower-reflectivity input mirror is because the lock to the cavity was more stable and the cavity-pole was much higher than the frequency region where we measured the EO effect.
We then tuned the $\lambda/2$ plate so that the laser polarization was aligned to the fast or slow axis where the EO effect can be observed.
As the birefingence in amorphous coatings are so small that the impact of replacing the input mirror is negligible~\cite{Bielsa2009}.

\subsection{Measurement scheme}

\begin{figure}[htbp]
\includegraphics[width=8.6cm]{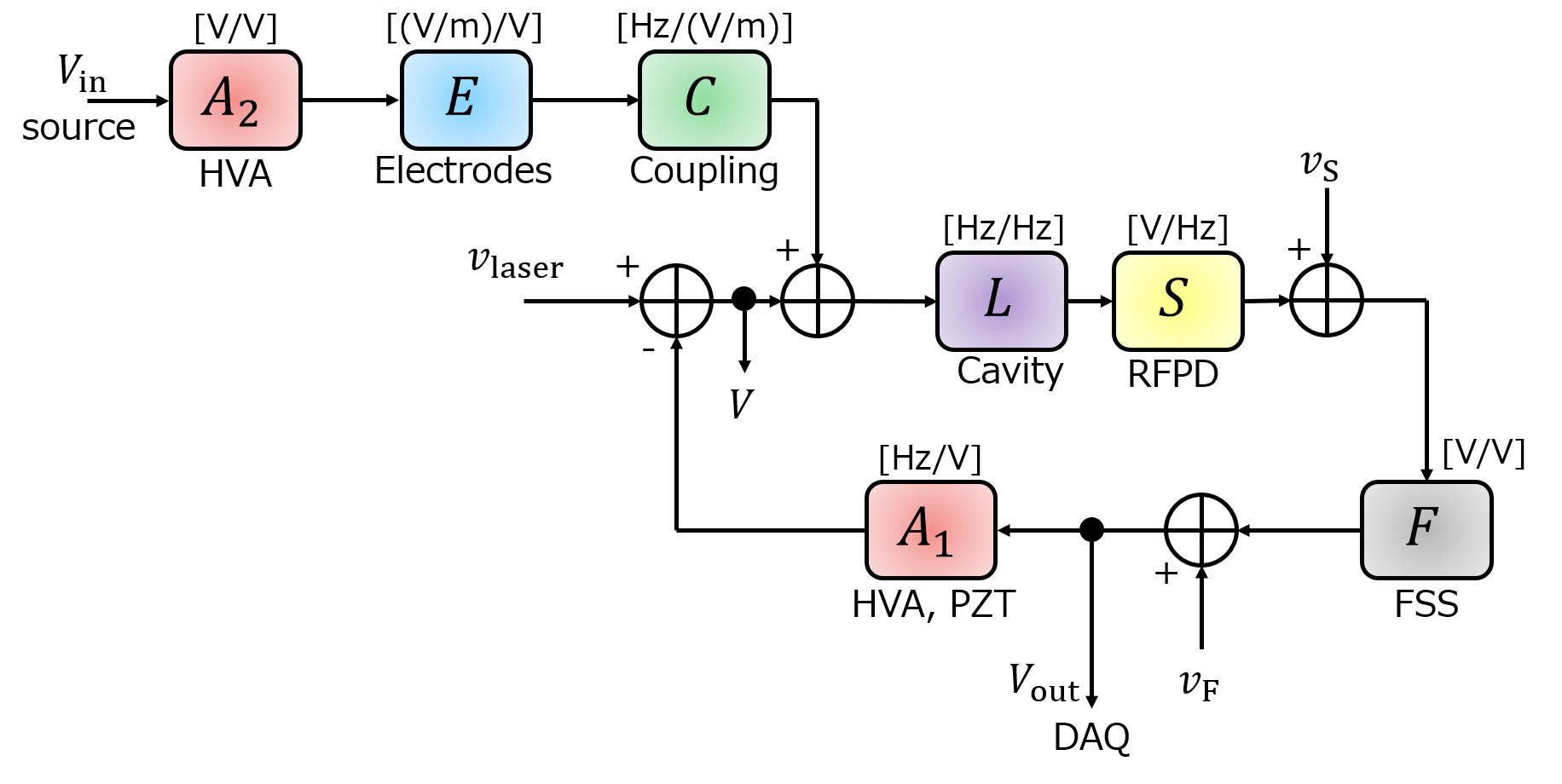}
\caption{Block diagram of measurement scheme. Transfer function from source signal $V_{\mathrm{in}}$ to PDH feedback signal $V_{\mathrm{out}}$ is measured by using a SR785. $v_{\mathrm{laser}}$, $v_{\mathrm{S}}$, and $v_{\mathrm{F}}$ denote the noises of the laser, RFPD, and FSS, respectively.}
\label{fig.scheme}
\end{figure}

Fig.~\ref{fig.scheme} shows the measurement scheme of our setup. 
In this scheme, fluctuations in the cavity displacement are probed by using the transfer function from the source signal, $V_{\mathrm{in}}$, to the feedback signal, $V_{\mathrm{out}}$.

The signal just after the summing node of the coupling from electric field, $V$, can be expressed as
\begin{align}
    V = \frac{1}{1+{G}}v_{\mathrm{laser}} - &\frac{{G}}{1+{G}}{CEA}_2V_{\mathrm{in}} \notag \\
    &- \frac{{FA}_1}{1+{G}}v_{\mathrm{S}} -  \frac{{A}_1}{1+{G}}v_{\mathrm{F}},
\end{align}
where $G \equiv A_1FSL$ is the open-loop gain.
The feedback signal becomes
\begin{align}
    V_{\mathrm{out}} &= {FSL}(V + CEA_2V_{\mathrm{in}}) + {F}v_{\mathrm{S}} + v_{\mathrm{F}}, \notag \\
    &= \frac{{FSL}}{1+{G}}CEA_2V_{\mathrm{in}} \notag \\ &\quad + \frac{{FSL}}{1+{G}}v_{\mathrm{laser}} + \frac{{F}}{1+{G}}v_{\mathrm{S}} + \frac{1}{1+{G}}v_{\mathrm{F}}.
\end{align}
Then, the transfer function, $V_{\mathrm{out}}/V_{\mathrm{in}}$, can be written as
\begin{align}
    \frac{V_{\mathrm{out}}}{V_{\mathrm{in}}} &= \frac{{FSL}}{1+{G}}{CEA}_2 + \frac{{FSL}}{1+{G}}\frac{v_{\mathrm{laser}}}{V_{\mathrm{in}}} \notag \\ &\quad + \frac{{F}}{1+{G}}\frac{v_{\mathrm{S}}}{V_{\mathrm{in}}} + \frac{1}{1+{G}}\frac{v_{\mathrm{F}}}{V_{\mathrm{in}}}.
    \label{eq.TF}
\end{align}
If the source signal $V_{\mathrm{in}}$ is much larger than the noises, $v_{\mathrm{laser}}$, $v_{\mathrm{S}}$, and $v_{\mathrm{F}}$, Eq. \ref{eq.TF} can be approximated as
\begin{align}
    \frac{V_{\mathrm{out}}}{V_{\mathrm{in}}} \approx \frac{FSL}{1+G}CEA_2 = \frac{G}{1+G}CE\frac{A_2}{A_1}.
    \label{eq.cal}
\end{align}
When $G$, $E$, $A_1$, and $A_2$ are known, coupling level of the EO effect, $C$, can be obtained from Eq.~($\ref{eq.cal}$).

\subsection{Calibration}

\subsubsection{Transfer function}

\begin{figure}[htbp]
\includegraphics[width=8.6cm]{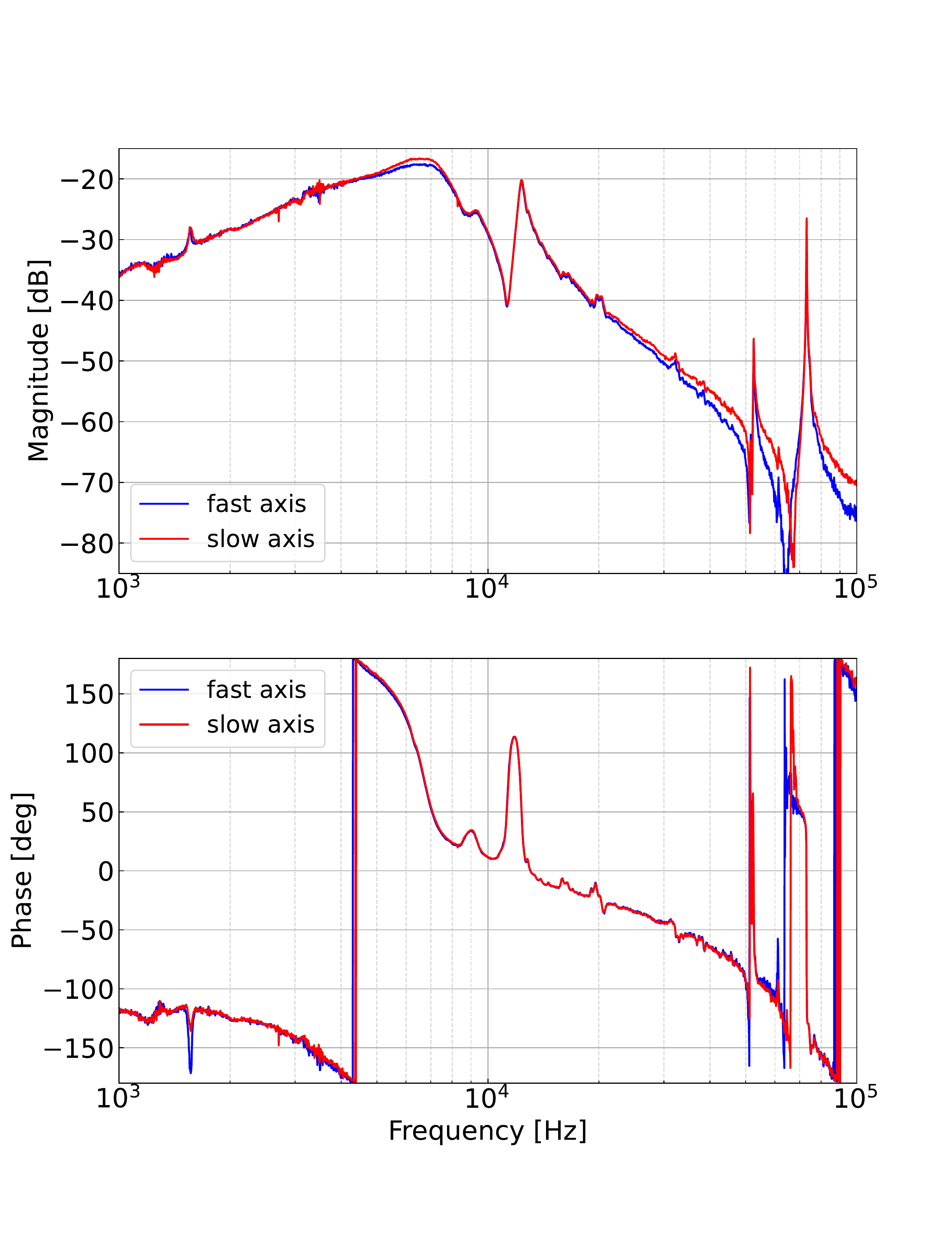}
\caption{
Measured transfer functions, $V_{\mathrm{out}}/V_{\mathrm{in}}$, for fast and slow axes.
}
\label{fig.rawTF}
\end{figure}

Fig.~\ref{fig.rawTF} shows the measured transfer functions, $V_{\mathrm{out}}/V_{\mathrm{in}}$, when the polarization is aligned to fast or slow axis.
As the unity gain frequency of PDH loop is about $4\unit{kHz}$, the fluctuations below $4\unit{kHz}$ are suppressed.
The electric field couples to the cavity length fluctuations through not only the EO effect, but also mechanical vibration.
Mechanical coupling through the mirror mount has a resonant peak around $10\unit{kHz}$.
Also, the peaks around $50\unit{kHz}$ and $70\unit{kHz}$ are mechanical resonances of the sample mirror.
Therefore, the measurements of the EO effect can be disturbed by the mechanical couplings in these frequency regions.


\subsubsection{Electric field}

\begin{figure}[htbp]
\includegraphics[width=8.6cm]{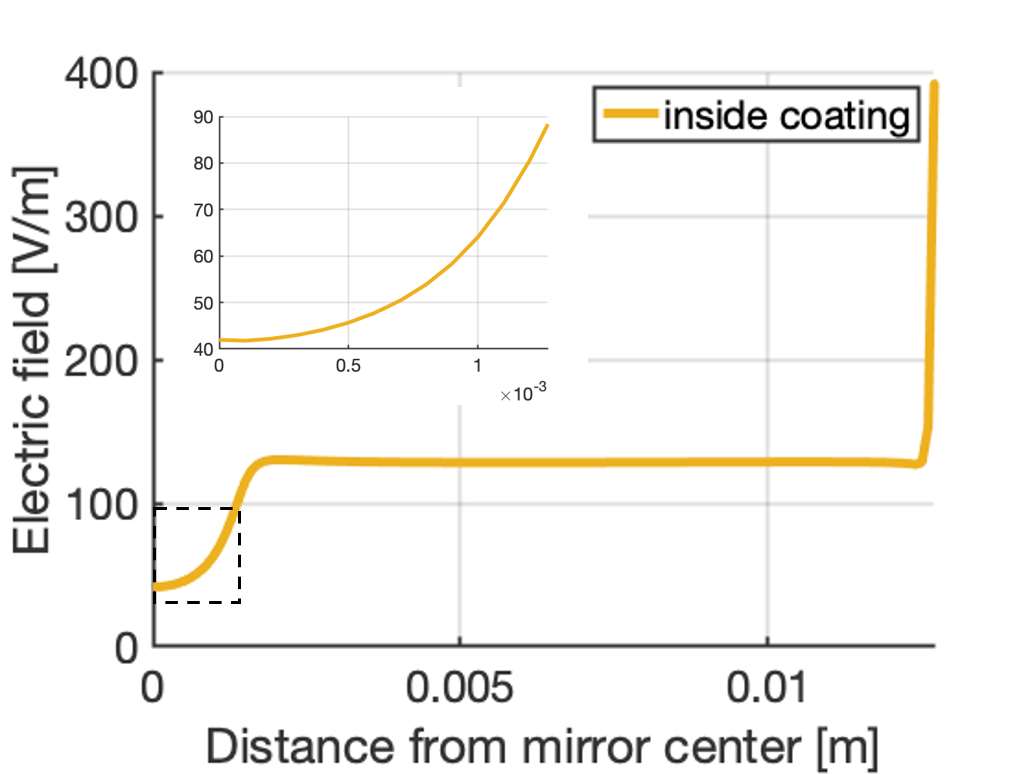}
\caption{
Calculated electric field normal to the AlGaAs coatings.
The horizontal axis is the distance from mirror center, and the vertical axis is the electric field $\mathrm{[V/m]}$ when the unit voltage is applied to the front electrode, i.e, the conversion efficiency, $E\unit{[(V/m)/V]}$.
}
\label{fig.efield}
\end{figure}

Voltage applied to the electrode is converted into an electric field which penetrates the AlGaAs sample mirror.
This conversion efficiency, $E\unit{[(V/m)/V]}$, is computed by an effective 3D static solution based on the geometry of the optics.
Fig.~\ref{fig.efield} shows the computed electric field when the unit voltage is applied to the front electrode i.e., the conversion efficiency, $E\unit{[(V/m)/V]}$.
The electric field close to the mirror center where the beam hits is $42\unit{V/m}$.
In our setup, the beam spot size on the AlGaAs mirror is about $100\unit{\mu m}$, and the electric field within the beam spot on the AlGaAs mirror can be treated as uniform.
Therefore, we apply the conversion efficiency as $E = 42\unit{(V/m)/V}$ and assume it is constant within the frequency region of interest.

\subsubsection{PZT response}

The internal PZT of the NPRO laser is used to actuate the laser frequency.
Its actuation efficiency is measured by scanning the laser frequency by a triangle wave.
Generally, the NPRO's laser PZT response has frequency dependence.
However, we cannot drive enough voltage to scan the laser frequency above a few kHz due to the low-pass filter of the HVA connected to the laser PZT.
On the other hand, the laser PZT response can be regarded as constant between $1 - 100\unit{kHz}$ \cite{klog}.
Therefore, we measured the actuation efficiency with $1\unit{kHz}$ triangle wave and assume that observed laser PZT efficiency is flat between $1 - 100 \unit{kHz}$.

\begin{figure}[htbp]
\includegraphics[width=8.6cm]{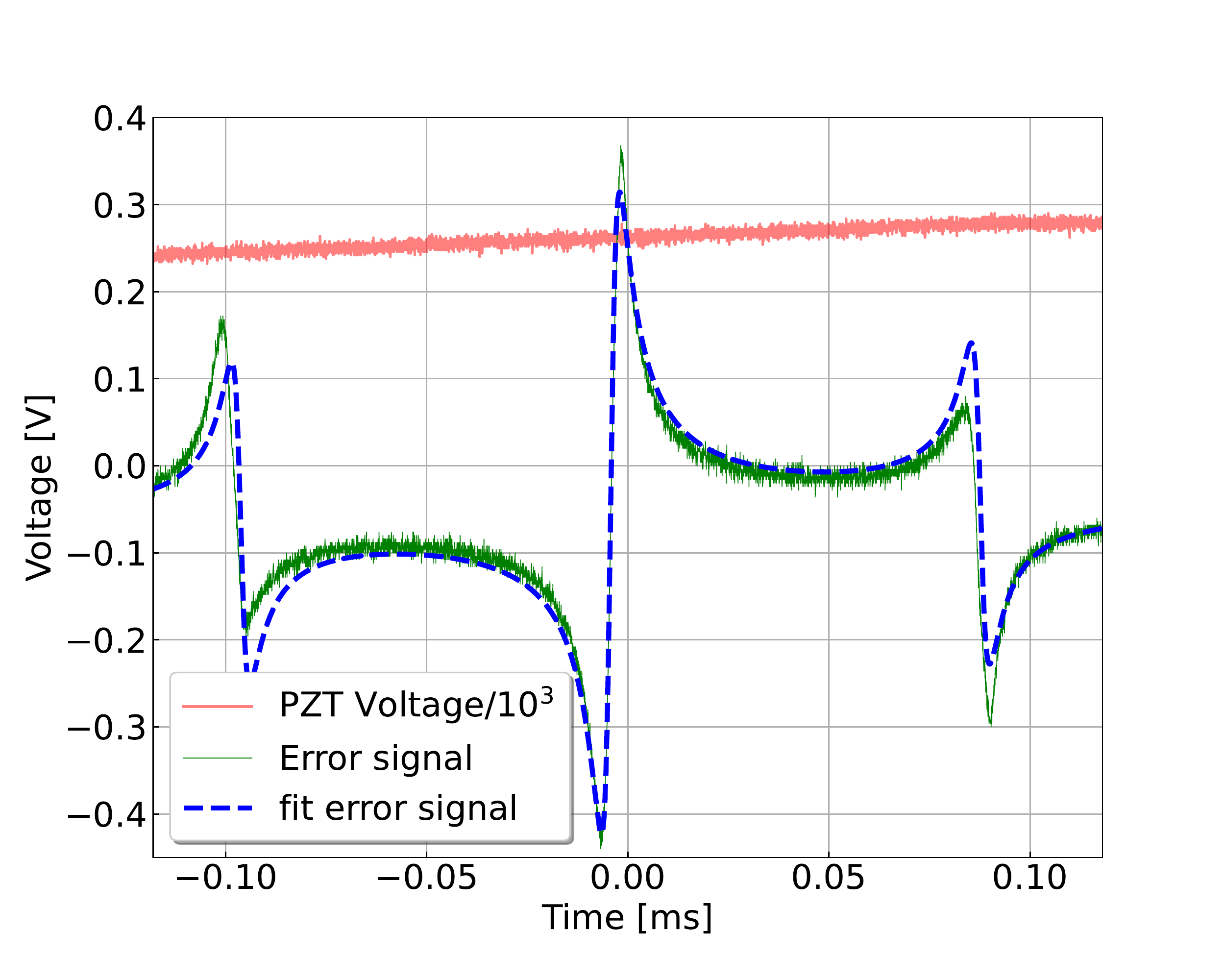}
\caption{
Error signal obtained by scanning the laser frequency with $1\unit{kHz}$ triangle wave.
Red and green solid lines correspond to the monitored voltage sent to NPRO's PZT and measured error signal, respectively.
Blue dashed line is fitted curve of error signal.
}
\label{fig.scan}
\end{figure}

Fig.~\ref{fig.scan} shows the response of the PDH error signal scanned by $1\unit{kHz}$ triangle wave.
We calculated the actuation efficiency by fitting the error signal.
From the fitting result, the PZT efficiency is estimated as $1.7\unit{MHz/V}$.

\section{Results}

\begin{figure}[htbp]
\includegraphics[width=8.6cm]{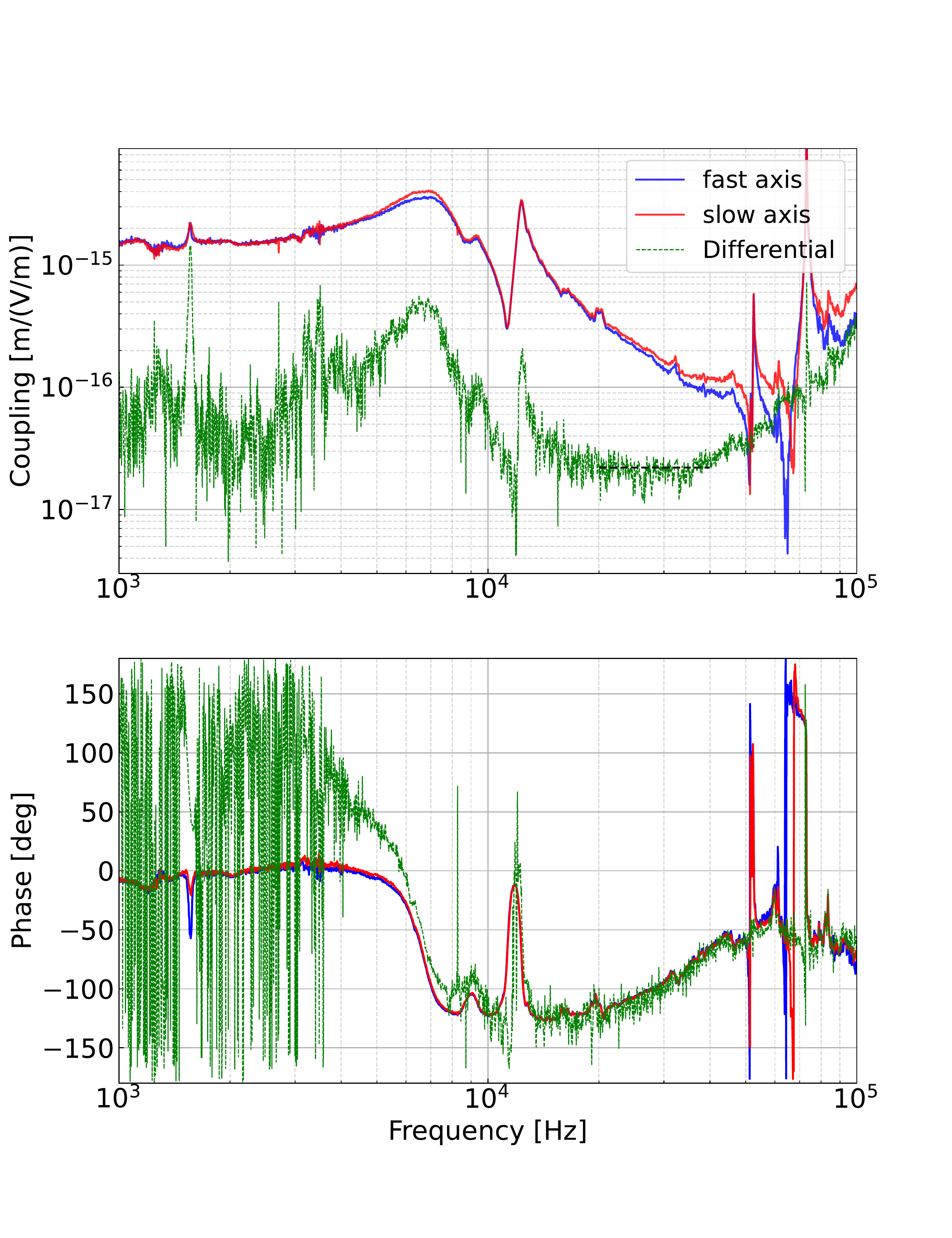}
\caption{
Calibrated transfer functions.
The black dashed line shows the fitted result between $20 - 40\unit{kHz}$.
}
\label{fig.result}
\end{figure}

From Eq.~(\ref{eq.cal}) and the obtained calibration data, one can evaluate the coupling between the electric field and cavity length, $C$.
Fig.~\ref{fig.result} shows the calibrated results of the coupling. 
Here we used Eq.~\ref{eq.cavity_fluctuation} to convert the unit from [Hz/(V/m)] to [m/(V/m)].
Measured coupling levels at both axes are almost the same at the order of $10^{-16}\unit{m/(V/m)}$.

The coupling, $C$, can be decomposed to mechanical coupling, $C_{\mathrm{m}}$, and the EO effect, $C_{\mathrm{EO}}$, as $C=C_{\mathrm{m}}+C_{\mathrm{EO}}$.
As described later, the signs of phase perturbation due to the EO effect are opposite between fast and slow axes.
This can be expressed as
\begin{align}
    C_{\mathrm{EO,\,slow}} &= |C_{\mathrm{EO}}|\mathrm{e}^{i\psi}, \\
    C_{\mathrm{EO,\,fast}} &= -|C_{\mathrm{EO}}|\mathrm{e}^{i\psi} \left(=|C_{\mathrm{EO}}|\mathrm{e}^{i(\psi+\pi)}\right),
\end{align}
where $\psi$ is the phase offset.
Assuming that the mechanical coupling is common to both fast and slow axes at this frequency region, differential between these two transfer functions becomes
\begin{align}
    \mathrm{Diff.} &= C_{\mathrm{slow}} - C_{\mathrm{fast}}  \notag \\
    &= \left(C_{\mathrm{m,\,slow}}+C_{\mathrm{EO,\,slow}}\right) - \left(C_{\mathrm{m,\,fast}}+C_{\mathrm{EO,\,fast}}\right) \notag \\
    &= 2|C_{\mathrm{EO}}|\mathrm{e}^{i\psi}.
\end{align}
Therefore, the magnitude of differential between the TFs of fast and slow axes ideally becomes twice the magnitude of the EO effect in AlGaAs coatings.

The green dashed line shown if Fig.~\ref{fig.result} is the differential between transfer functions of fast and slow axes, $2|C_{\mathrm{EO}}|\mathrm{e}^{i\psi}$.
Mechanical couplings from the mirror mount and resonances of mirror itself disturb the cavity length below $\sim10{\unit{kHz}}$ and around $\sim50 - 80\unit{kHz}$.
Above $\sim40\unit{kHz}$, the differential shows the frequency dependence.
One possibility of this behavior is that the frequency dispersion of the electro-optic coefficients of GaAs and AlGaAs~\cite{Abarkan2003}.
Further studies may be needed to fully understand the behavior at those higher frequency region.
On the other hand, for the case of current terrestrial GWDs, the important frequency region is between $\sim10\unit{Hz}$ and several kHz.
As shown in the previous work, the electro-optic coefficient tend to show the flat response below a few tens of kHz~\cite{Abarkan2003}.
We focus on the frequency region $20-40\unit{kHz}$ where the impacts of mechanical couplings can be considered small and the differential has flat response, and assume that the EO effect in AlGaAs coatings is frequency independent below $40\unit{kHz}$.
From the above assumptions, we obtain $2|C_{\mathrm{EO}}|= 2.2\times10^{-17}\unit{m/(V/m)}$ by fitting the result.
Therefore, the coupling level of the EO effect is estimated as $|C_{\mathrm{EO}}|=1.1\times10^{-17}\unit{m/(V/m)}$.

\section{Discussions}

\subsection{Comparison to theoretical estimation}

The level of EO effect can be numerically computed by using a transfer matrix calculation of the coating multilayer.
The perturbation of the reflected field phase induced by $k$-th coating layer can be described as \cite{Evans2008, Ballmer2015}
\begin{align}
    \frac{\partial\phi_{\mathrm{c}}}{\partial\phi_k} = \Im \left(\frac{1}{M_{21}}\frac{\partial M_{21}}{\partial\phi_k} - \frac{1}{M_{22}}\frac{\partial M_{22}}{\partial\phi_k}\right),
\end{align}
where $M_{ij}$ are elements of the transfer matrix of coatings, $M$, and $\Im$ denotes the imaginary part.
The transfer matrix of the total coating can be given by
\begin{align}
    M = Q_ND_N \cdots Q_kD_k \cdots Q_1D_1Q_0,
    \label{eq.TFmatrix}
\end{align}
where $Q_0$ is the transition between vacuum and 1st layer, and $Q_k$ is the transition matrix from $k$-th layer to $(k+1)$-th layer defined as
\begin{align}
    Q_k = \frac{1}{2n_{k+1}}
            \begin{pmatrix}
            n_{k+1}+n_k & n_{k+1}-n_k \\
            n_{k+1}-n_k & n_{k+1}+n_k 
            \end{pmatrix}.
    \label{eq.transmat}
\end{align}
$D_k$ is the propagator through the $k$-th coating layer expressed as
\begin{align}
    D_k = \begin{pmatrix}
            \mathrm{e}^{-i\phi_k/2} & 0 \\
            0 & \mathrm{e}^{i\phi_k/2}
            \end{pmatrix},
    \label{eq.propmat}
\end{align}
where $\phi_k=4\pi n_kd_k/\lambda$ is round trip phase change.
From Eqs.~(\ref{eq.TFmatrix}) $-$ (\ref{eq.propmat}), partial derivative of transfer matrix can be calculated as
\begin{align}
    \frac{\partial M}{\partial\phi_k} = Q_ND_N \cdots Q_k&D_k
    \begin{pmatrix}
    -i/2 & 0 \\
    0 & i/2
    \end{pmatrix}
    \notag \\
    &Q_{k-1}D_{k-1} \cdots Q_1D_1Q_0.
\end{align}
From the definition of round trip phase change, $\phi$, $\partial\phi_k/\partial E$ becomes
\begin{align}
\frac{\partial\phi_k}{\partial E} = \frac{4\pi d_k}{\lambda}\frac{\partial n_k}{\partial E} = \pm\frac{2\pi}{\lambda}n_k^3d_kr_{41,k},
\end{align}
where the signs depend on the AlGaAs axes.
By using the chain rule, the phase perturbation induced by the electro-optic effect can be expressed as
\begin{align}
    \frac{\partial\phi_{\mathrm{c}}}{\partial E} = \frac{\partial\phi_{\mathrm{c}}}{\partial\phi_k}\frac{\partial\phi_k}{\partial E}.
\end{align}
Here we assume the EO coefficients of $\mathrm{GaAs}$ and $\mathrm{Al}_{x}\mathrm{Ga}_{1-x}\mathrm{As}$ as $r_{41,\mathrm{GaAs}}=-1.33\times10^{-12}\unit{m/V}$, and $r_{41,\mathrm{AlGaAs}}=-(1.33-0.45x)\times10^{-12}\unit{m/V}$, respectively~\cite{Adachi1985, Abernathy2014}.
As a result, one can compute the phase perturbation induced by the electro-optic effect as
\begin{align}
    \left|\frac{\partial\phi_{\mathrm{c}}}{\partial E}\right| = \left|\frac{\partial\phi_{\mathrm{c}}}{\partial\phi_k}\frac{\partial\phi_k}{\partial E}\right| = 3.9 \times 10^{-11}\unit{rad/(V/m)}.
\end{align}
This phase perturbation can be converted to the Fabry-Perot cavity displacement, $\partial L/\partial E$.
Round trip phase of a Fabry-Perot cavity, $\phi$, satisfies the relationship as
\begin{align}
    \phi = \frac{2L\omega}{c} = \frac{4\pi L}{\lambda},
    \label{eq.cav}
\end{align}
where $L$, $\omega$, $c$, and $\lambda$ are the cavity length, angular frequency, the speed of light, and the wavelength of laser, respectively.
From Eq.~(\ref{eq.cav}), one can obtain
\begin{align}
    \frac{\partial\phi}{\partial E} &= \frac{4\pi}{\lambda}\frac{\partial L}{\partial E}.
\end{align}
Consequently, the coupling of EO effect to cavity length can be calculated as
\begin{align}
    \left|\frac{\partial L}{\partial E}\right| = \frac{\lambda}{4\pi}\left|\frac{\partial\phi_{\mathrm{c}}}{\partial E}\right| = 3.3\times10^{-18}\unit{m/(V/m)}.
\end{align}
This value is about one-third of the measured value.

\subsection{Implications for gravitational wave detectors}

We evaluate the impacts of noise induced by the EO effect on future GWDs.
In GWDs such as aLIGO, horizontally polarized (P-polarized) beam is employed for laser interferometry~\cite{Aasi2015}.
The impacts of the EO effect on GWDs depend on the alignment between the beam polarization and AlGaAs axes.

Firstly, we consider the case that the polarization of the beam is aligned to the AlGaAs $[110]$ or $[1\bar{1}0]$ axes ($x'$ or $y'$) where the reflected optical phase is perturbed by the EO effect.
The measured fluctuations in the electric field next to the test mass in aLIGO is $3\times10^{-6}\unit{(V/m)/\sqrt{Hz}}$ at $100\unit{Hz}$~\cite{Buikema2020}.
Assuming that the fluctuations in the electric fields next to each of the four test masses are the same level, and uncorrelated with each other, the strain noise due to the EO effect at $100\unit{Hz}$ can be calculated as
\begin{align}
    \frac{\sqrt{4}\times1.1\times10^{-18}\unit{m/(V/m)}\times3\times10^{-6}\unit{(V/m)/\sqrt{Hz}}}{4\times10^3\unit{m}} \notag \\
    = 1.6\times10^{-26}\unit{1/\sqrt{Hz}}.
\end{align}
Here we assumed that the EO effect has flat response and the arm cavity length is $4\unit{km}$.
The target sensitivity of A+, future upgrade plan of aLIGO, is about $2\times10^{-24}\unit{1/{\sqrt{Hz}}}$ at $100\unit{Hz}$~\cite{Barsotti2018a+}.
Therefore, the noise level of EO effect is about two orders of magnitude smaller than the sensitivity of A+.
As long as fluctuations in the electric field are kept below $\sim2\times10^{-5}\unit{(V/m)/\sqrt{Hz}}$ at $100\unit{Hz}$, the noise level of the EO effect is below $10^{-25}\unit{1/\sqrt{Hz}}$, and will not affect the sensitivity of GWDs.

\begin{figure}[htbp]
\includegraphics[width=8.6cm]{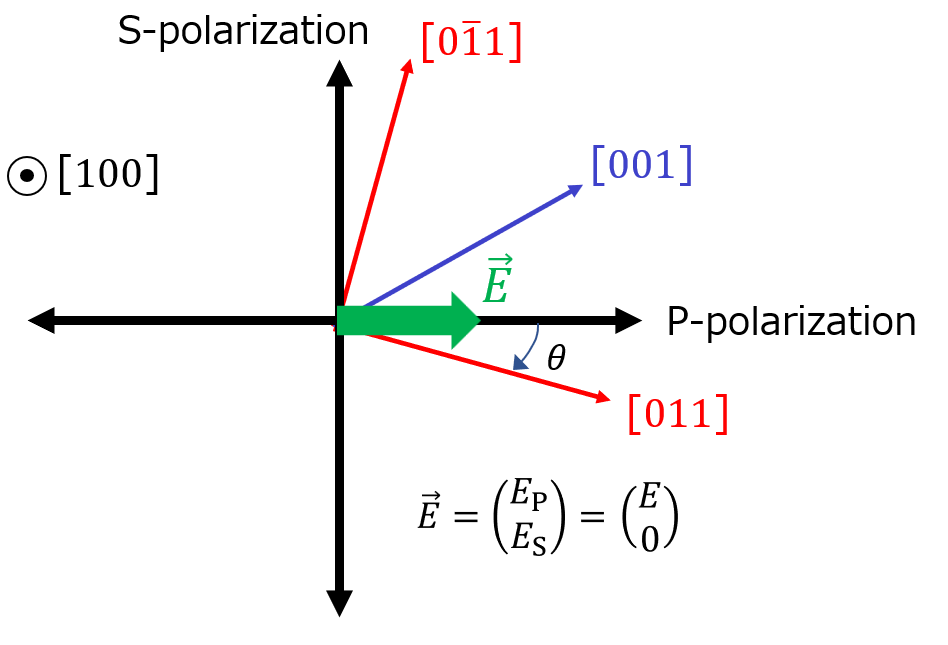}
\caption{
Schematic of the electric field of horizontally polarized beam, $\vec{E}$, and tilted AlGaAs axes.
}
\label{fig.tilt}
\end{figure}

Secondly, we consider the case when the $x'$ and $y'$ axes are tilted $\theta$ degrees from the beam polarization as shown in Fig.~\ref{fig.tilt}.
By defining the field of the beam as $\begin{pmatrix}E_0 & 0\end{pmatrix}^{\mathrm{T}}$, its projection onto $[011]$ and $[0\bar{1}1]$ axes can be expressed as
\begin{align}
    \begin{pmatrix}
    E_{[011]} \\ E_{[0\bar{1}1]} \\
    \end{pmatrix}
    =
    \begin{pmatrix}
    \cos\theta & -\sin\theta  \\ \sin\theta & \cos\theta \\
    \end{pmatrix}
    \begin{pmatrix}
    E_0 \\ 0 \\
    \end{pmatrix}
    =
    E_0
    \begin{pmatrix}
    \cos\theta \\ \sin\theta \\
    \end{pmatrix}.
\end{align}
We denote the optical phase perturbation induced by the EO effect as $\phi_{\mathrm{EO}}$.
Then the field perturbed by the EO effect becomes
\begin{align}
    \begin{pmatrix}
    \tilde{E}_{[011]} \\ \tilde{E}_{[0\bar{1}1]} \\
    \end{pmatrix}
    &=
    E_0
    \begin{pmatrix}
    \mathrm{e}^{i\phi_{\mathrm{EO}}} & 0 \\ 0 & \mathrm{e}^{-i\phi_{\mathrm{EO}}} \\
    \end{pmatrix}
    \begin{pmatrix}
    \cos\theta \\ \sin\theta \\
    \end{pmatrix}.
\end{align}
By converting the coordinates from AlGaAs axes to beam polarization axes, one can get
\begin{align}
    \begin{pmatrix}
    E_{\mathrm{P}} \\ E_{\mathrm{S}} \\
    \end{pmatrix}
    &=
    \begin{pmatrix}
    \cos\theta & \sin\theta  \\ -\sin\theta & \cos\theta \\
    \end{pmatrix}
    \begin{pmatrix}
    \tilde{E}_{[011]} \\ \tilde{E}_{[0\bar{1}1]} \\
    \end{pmatrix}, \notag \\
    &= E_0
    \begin{pmatrix}
    \mathrm{e}^{i\phi_{\mathrm{EO}}}\cos^2\theta + \mathrm{e}^{-i\phi_{\mathrm{EO}}}\sin^2\theta \\
    -\mathrm{e}^{i\phi_{\mathrm{EO}}}\cos\theta\sin\theta + \mathrm{e}^{-i\phi_{\mathrm{EO}}}\cos\theta\sin\theta \\
    \end{pmatrix}.
    \label{eq.field}
\end{align}
Assuming $|\phi_{\mathrm{EO}}|\ll1$, Eq.~(\ref{eq.field}) can be approximated as
\begin{align}
    \begin{pmatrix}
    E_{\mathrm{P}} \\ E_{\mathrm{S}} \\
    \end{pmatrix}
    &\approx E_0
    \begin{pmatrix}
    1 + i\phi_{\mathrm{EO}}(\cos^2\theta - \sin^2\theta) \\ -2i\phi_{\mathrm{EO}}\cos\theta\sin\theta \\
    \end{pmatrix}.
    \label{eq.biref}
\end{align}
When the polarization of the beam is aligned to the $[100]$ axis, $\theta=45\unit{deg}$, Eq.~(\ref{eq.biref}) becomes
\begin{align}
    \begin{pmatrix}
    E_{\mathrm{P}} \\ E_{\mathrm{S}} \\
    \end{pmatrix}
    &\approx E_0
    \begin{pmatrix}
    1  \\ -i\phi_{\mathrm{EO}} \\
    \end{pmatrix}.
\end{align}
Therefore, the EO effect in AlGaAs coatings induces birefringence, and a P-polarized beam is generated by this effect, leading to elliptical polarization.
However, the amplitude of the P-polarized beam converted from S-polarization is the order of $|\phi_{\mathrm{EO}}|\sim10^{-16}$, and it can be negligible.
Moreover, the reflected phase of the S-polarized beam is not disturbed by the EO effect.
As a result, in the ideal case, the impacts of the EO effect can be mitigated when the polarization is aligned to the $[001]$ family of axes.
However, those axes can show the static birefringence as shown in Fig.~\ref{fig.split}.
From Eq.~(\ref{eq.cav}) and the laser frequency scan measurement, the birefringence in our AlGaAs coating is $\Delta\theta_{\mathrm{b}}\approx2.2\times10^{-3}\unit{rad}$.
Considering the case of a km-scale GWD such as aLIGO, the resonant frequency split of arm cavity becomes
\begin{align}
    \Delta\nu &= \frac{c}{4\pi L}2\Delta\theta_{\mathrm{b}} \notag \\
    &\approx 26\unit{Hz}.
\end{align}
Here we assumed that the arm cavity length $L$ is $4\unit{km}$ and the both input and end mirrors of the arm cavity have the same amount of birefringence, $\Delta\theta_{\mathrm{b}}=2.2\times10^{-3}\unit{rad}$.
Two orthogonal eigenmodes generated by the static birefringence in AlGaAs coatings will be within the FWHM of arm cavity ($\sim80\unit{Hz}$), and have a potential to interfere with control loops that have a similar bandwidth.



Even when the beam polarization is aligned to the AlGaAs $[011]$ family of axes where the EO effect is maximized, the noise induced by the EO effect is well below the design sensitivity.
Moreover, when one of the $[001]$ family of axes is aligned to the beam polarization, further reduction in the EO effect could be realized without serious birefringence.
As a result, the EO effect in AlGaAs coatings will not be a limiting noise source in future GWDs, and employing AlGaAs test masses will enhance the scientific outcomes which can be obtained from observations.

It should be noted that further studies are needed to realize AlGaAs coating test masses in future GWDs.
The beam spot size on the test masses will be larger than $5\unit{cm}$.
In order to avoid clipping losses of the beam, the coating will need to be scaled to $30\unit{cm}$ diameter for current detectors and to about $50\unit{cm}$ for next generation detectors.
Moreover, AlGaAs coatings are opaque to $532\unit{nm}$ laser light, which is currently used in GWDs for cavity-length stabilization~\cite{Izumi2012, Staley2014}.
Thus, a new scheme with a transparent stabilization beam will need to be defined.
Finally, the impact of the birefringence of AlGaAs coatings on GWDs with a focus on identifying the root cause of this effect must continue to be investigated~\cite{Winkler2021, Michimura2022, Yu2022}.
We suggest this research include exploring alternative orientations of the crystalline structure that may minimize this effect.

\section{Conclusion}

Crystalline AlGaAs coatings, with their lower coating thermal noise, have the potential to dramatically  improve the sensitivity and detection rate of GWDs and greatly bolster the new field of GW astrophysics
We investigate the noise induced by the EO effect in AlGaAs coating caused by the fluctuations in the electric field.
This study yields that the EO effect will not be a limiting noise source in future upgraded GWDs.

Our study helps pave a path for utilizing AlGaAs mirror coatings in future upgraded GWDs.
Further studies will lead to the large-area substrate transferred crystalline test mass coatings.

\begin{acknowledgments}
This work was supported with funding from the National Science Foundation grants:  PHY-1707863, PHY-1912699, PHY-2011688, and PHY-2011723. 
A portion of this work was performed in the UCSB Nanofabrication Facility, an open access laboratory.
This paper has LIGO Document number LIGO-P2200244.

\end{acknowledgments}



\nocite{*}

\bibliography{apssamp}

\end{document}